\documentclass[12pt,preprint]{aastex}

\shorttitle{RESONANT TEMPERATURE FLUCTUATIONS IN NEBULAE}
\shortauthors{Bautista \& Ahmed}
\begin{document}
\title{RESONANT TEMPERATURE FLUCTUATIONS IN NEBULAE IONIZED BY SHORT-PERIOD BINARY STARS}

\author{Manuel A. Bautista and Ehab E. Ahmed} 
\affil{Department of Physics, Western Michigan University,
Kalamazoo, MI 49008, USA}
\email{manuel.bautista@wmich.edu}

\begin{abstract} 
A prevailing open problem in planetary nebulae research, and photoionized gaseous nebulae research at large,
is the systematic discrepancies in electron temperatures and ionic abundances as derived from
recombination and collisionally excited lines.
Peimbert (1967) proposed the presence of 'temperature fluctuations' in
these nebulae, 
but the apparent amplitude of such fluctuations, as deduced from spectral
diagnostics and/or abundance discrepancy factors, remain unexplained by standard 
photoionization modeling.
While this and other alternative models to explain the temperature and
abundance discrepancies remain 
inconclusive, recent observations seem to point at a connection between
nebular abundance discrepancy factors and a binary nature of photoionizing 
stars.
In this paper we show that large amplitude temperature fluctuations are expected to form in planetary nebulae photoionized by
short-period binary stars. 
Resonant temperature fluctuations are first formed along the orbital disk
around the binary stars, as the periodically varying ionizing radiation field
induces
periodic oscillations in the heating-minus-cooling function. 
Then, the temperatures fluctuations propagate vertically to the disk as 
thermal waves that later steepen into radiative shocks.
The binary period of the ionizing stars is determinant in the 
formation and propagation of temperature fluctuations, as well as in associated
density fluctuations. Fluctuations propagate efficiently only in systems with 
binary periods significantly shorter
than the gas thermalization time, of the order of 10 days.

Further, we propose temperature diagnostic line ratios that combine [\ion{O}{3}] 
collisionally excited lines and \ion{O}{2}
recombination lines to determine the equilibrium temperature and 
the magnitude of resonant temperature fluctuations in nebulae.
\end{abstract}


\normalsize

\section{INTRODUCTION}

Arguably, the most intriguing question left unanswered in photoionization modeling in
astronomy pertains the origin of systematic discrepancies in ionic abundances derived from
recombination and collisionally excited lines in a large fraction of known 
\ion{H}{2} regions and planetary nebulae (PNe).
Such differences in derived abundances 
are generally quantified in terms of abundance discrepancy factors (ADF)
that can reach up to two orders of magnitude for C, N, O, and Ne in
some extreme PNe (e.g. McNabb et al. 2016; Corradi et al. 2015;
Wesson et al. 2003). These ADF seem
to be the result of temperatures associated with recombination spectra being
considerably lower than those derived from forbidden collisionaly excited lines
(Torres-Peimbert et al. 1980). Peimbert (1967) proposed the existence of 
``temperature fluctuations" common to \ion{H}{2} regions and PNe, but the amplitude of such fluctuations needed to
reconcile the abundance determination are too large, in general, to be reproduced by standard
photoionization modeling (Kingdon and Ferland 1995). The existence of 
temperature variations of some sort has been
supported by modern spectra from high sensitivity, high spatial resolution instruments. 
Liu et al. (2000, 2001) showed that the temperatures of PNe
determined by ratios of collisionally excited lines (e.g. [\ion{O}{3}], 
[\ion{N}{2}], [\ion{S}{3}]) are typically larger
than the temperatures derived in hydrogen by fitting the Balmer discontinuity to the Balmer 
recombination lines (Te(Bac)). Moreover, the
ADF from collisional and recombination lines from optical and UV
spectra of PNe and \ion{H}{2} regions are correlated with the 
difference between Te([\ion{O}{3}])
and Te(Bac) (Garc\'{\i}a-Rojas and Esteban 2007; Liu et al. 2004). Further, 
point-to-point electron temperature variations 
have been obtained for several high surface brightness PNe and 
\ion{H}{2} regions (Rubin et al. 2002, 2003;
Krabbe and Copetti 2002, 2005; O'Dell et al. 2003, 2013; Garnett and Dinerstein 2001; Wesson \& Liu 2004),
though the spatial scale of the variations is expected to be too small to be resolved in detail.
Not surprisingly, temperature variations averaged over the smallest spatial 
scales that can be resolved observationally at present are too small to account
for ADF (see the case of NGC~6543 by Wesson \& Liu 2004).

At present, the idea that is receiving
the most attention in explaining ADF is the hypothesis of 
chemically inhomogeneities. According to this hypothesis there would be in the nebula pockets 
of cold very metal-rich plasma mixed with the gas or an extended high metallicity gas embedded
in a less dense ambient gas with lower metallicity (Tsamis et al. 2003).
The former idea was first suggested by Torres-Peimbert et al. (1990) and has
been 
studied most extensively by Liu et al. (2006). 
Though, this idea lacks workable models that explain the origin of such
chemical inhomogeneities.        
Henney \& Stasi\'nska (2010) tried to 
explain the presence of metal-rich droplets in PNe by destruction of
solid bodies; 
however, they concluded that the amount of solid bodies needed to reproduce the observations was anomaly large.

In recent years, it has been found that a large fraction of all
intermediate-mass stars, PNe progenitors are known to be in binary systems 
(Moe and De~Marco 2006; Miszalski et al. 2009). Moreover,
there is now mounting observational 
evidence that binary stars play a significant role
in the PN ejection process. In particular, it seems like all PNe with extreme abundance
discrepancy factors host short-period binary stars (Corradi et al. 2015; Wesson et al. 2017).
Observational searches for close binaries demonstrate
that these are virtually all found in bipolar PNe (Mastrodemos and Morris 1999; Miszalski et al. 2009). Though, because searches for binary stars
are mostly limited to near edge-on systems with significant photometric variability it remains
unproven whether all non-spherical PNe, which in fact are the large majority of PNe, are ejected
from a star in a binary system. Surveys have found that between $\sim$10\% and 20\% of all PNe central stars are
close binary systems, with periods typically shorter than 3 days (Jones et al. 2016). 

Here we show that the periodically varying photoionizing radiation field
of a short-period binary star will lead to resonant temperature fluctuations (RTF) in
the orbital disk and these can propagate through the rest of the cloud
as thermal waves that lead to radiative shocks. 
The mechanisms for this process are described in the
next section. Further, Section 3 presents spectral diagnostics to 
determine the equilibrium and resonant temperature fluctuation amplitude from
observed spectra.

\section{Resonant Temperature Fluctuations (RTF)} 

In this section we present a theoretical model to explain RTF in PNe photoionized by binary stellar systems.
To this end, we first look at the formation of resonant fluctuations along the
orbital disk of the binary. Later, we study how these fluctuations may
propagate perpendicularly to the disk to the rest of the nebula.

In order to have a semi-realistic model we look at the geometry of known close-binaries.
UU~Sagittae is the central eclipsing binary system of the planetary nebula Abell
 63 (Bond et al. 1978). The total mass of the system is between 1
 and 3 $M_{\sun}$ and the orbital separation is in
the range of 2.5 - 3.6 $R_{\sun}$. The effective temperatures of the stars are $\sim
$35000~K for the primary and $\sim$4600~K for the secondary.
This means that only the primary star is able to radiate UV ionizing photons and this flux is mostly shut off during the eclipse by the
secondary star. The orbital period for this system is 0.465 days,
with a total eclipse duration of about one tenth of the period.
This also indicates that the thickness of the orbital disk is of about 15$^o$.

\subsection{Fluctuations in the orbital plane}

Let us assume a stationary, inviscid and nonconducting medium in plane parallel symmetry. The equations of conservation of momentum and energy are
\begin{equation}
\frac{\partial u}{\partial t}= \frac{1}{m n}\frac{\partial P}{\partial x}
\end{equation}
and
\begin{equation}
\frac{\partial \epsilon}{\partial t} = \frac{P}{n}\frac{\partial u}{\partial x} +
L.
\end{equation}
Here, $u$ is the velocity, $n$ is the total atomic number density, $m$ is the average mass per particle,  $P$ is the pressure, $\epsilon$ is the internal energy per particle,
$L$ is the heating-minus-cooling rate.
For a perfect gas dominated by thermal pressure we have 
\begin{equation}
P = 2\times n k_B T + P_0
\end{equation}
and
\begin{equation}
\epsilon = 2\times (3/2) k_B T,
\end{equation}
where $T$ is the temperature and $P_0$ includes all forms of pressure other than thermal pressure. 
The factor of 2 in both relationships comes from the fact that in a plasma where hydrogen is mostly ionized the number density of particles in the plasma is about twice the density of atoms.

Let us introduce a small temperature perturbation and assume that the density is constant through the cloud 
and in time, so that 
\begin{equation}
T = T_0(x) + T_1(x,t),  
\end{equation}
where $T_0$ is the steady state equilibrium temperature and $T_1$ is the time dependent perturbation.
From equations (1) through (5) we have the temperature fluctuations equation  
\begin{equation}
\frac{\partial^2 T_1(x,t)}{\partial t^2} = v_s^2\frac{\partial^2 T_1(x,t)}{\partial x^2}
+ \frac{1}{k_B} \frac{\partial L}{\partial t}, 
\end{equation}\label{waveeqn}
where, $v_s=\sqrt{(2/3)(k_B T_0/m)}$ is the speed of sound.    

In steady state conditions $L$ is identically zero for $T=T_0$ and 
$\frac{\partial L}{\partial T}|_{T=T_0} < 0.$
Hence, this term will damp any temperature fluctuations in the nebula. 
On the other hand, if the nebular ionizing source varies with time so will the heating and cooling rates. 
Along the orbital disk of the binary system the ionizing flux will vary as the
hotter star is periodically eclipsed by the cooler one.

For any ionizing binary system the 
heating-minus-cooling function for gas in the orbital disk will vary periodically and can always be represented 
by a combination of oscillation modes of the form
\begin{equation}
L(t,x)= \sum_j L_j \exp{[i(\omega t - \kappa_j x)]}.
\end{equation}

\begin{figure}
\rotatebox{0}{\resizebox{\hsize}{\hsize}
{\plotone{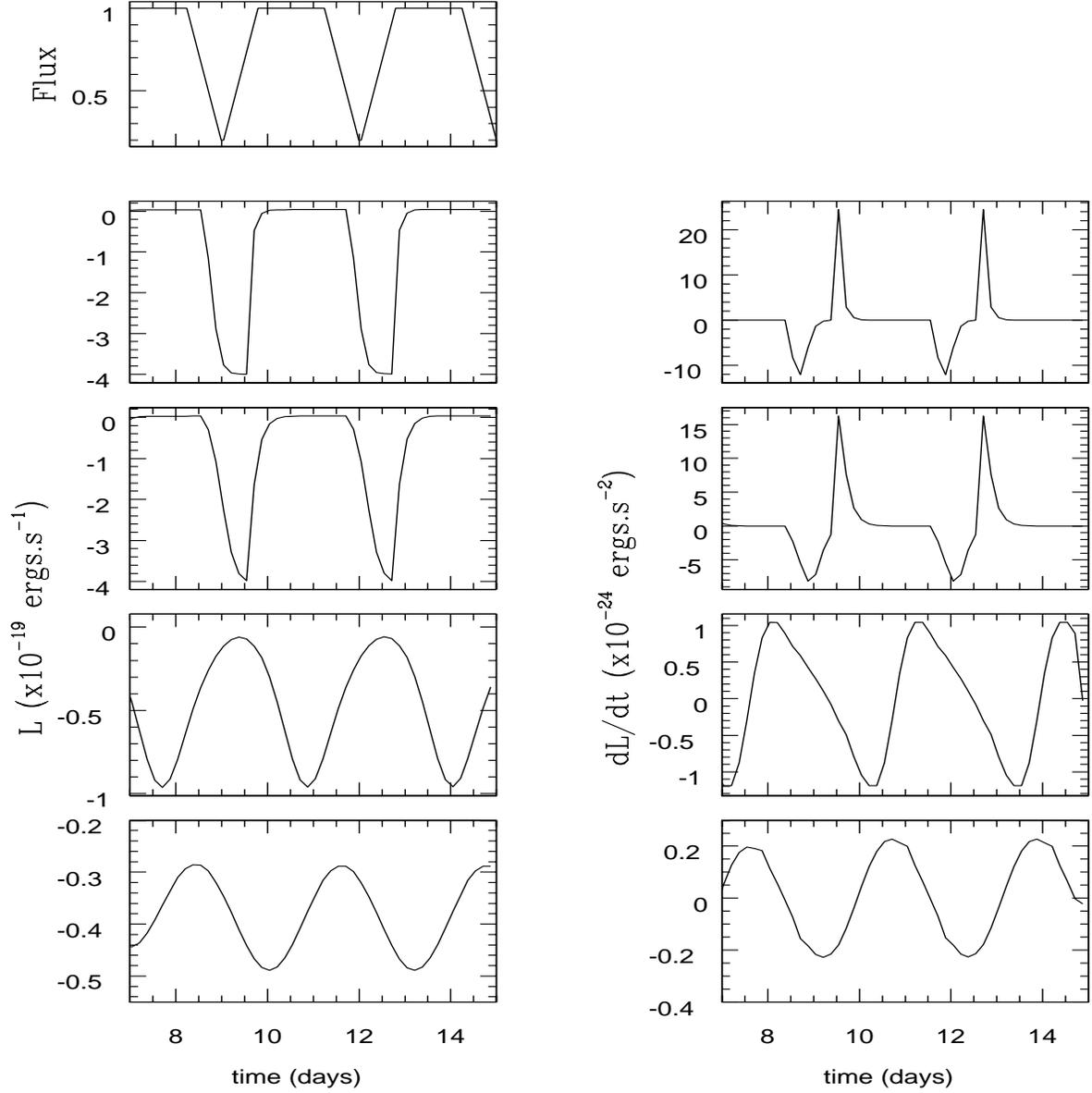}}}
\caption{Light curve of the ionizing source (top panel), heating-minus-cooling rates (four left panels), and time derivatives 
of the heating-minus-cooling rates (right panels). The heating-minus-cooling rates are shown for 
radii $2.3\times 10^{14}$, $6.2\times 10^{14}$, 
$2.9\times 10^{15}$, and $7.5\times 
10^{15}$~cm. 
}
\label{hmcfunction}
\end{figure} 
 
Figure 1 shows the results of a simulation for a binary star with occultation period of 3 days. 
For this simulation we pick a period of 3~days as representative of short-period binaries. As we will show here, shorter periods
as in UU~Sagittae, would lead to larger temperatures fluctuations. The top panel of Figure~1 
depicts the normalized light curve of the ionizing source.
The lower panels show the heating-minus-cooling function, $L$,
and its derivative versus time for four different depths within the 
cloud. 
For this model we assume an spectral energy distribution given by a black body at 50,000~K and 
an ionization parameter at the spectral maximum $\log{(\xi)}=-1.0$.
The
circumstellar nebula has a constant density $n_H=10^4$~cm$^{-3}$ and 
a chemical mixture containing H, He, C, N, O, S, and Fe with solar abundances.
We assume that the ionized nebula is radiation bound, which for the present
ionization parameter would have an extension of $\sim 3\times 10^{17}$~cm
up to the ionization front. For this nebula, $O^{++}$ is the dominant stage
of oxygen in the range from $\sim 10^{14}$~cm to $\sim 10^{17}$~cm. 
Clearly, the qualitative results are the same for nebulae with 
different ionization parameters, densities, and compositions.
The simulation was done with the time-dependent version of the photoionization
modeling code XSTAR (Ahmed 2017; Garc\'{\i}a et al. 2013; Kallman and Bautista 1999).

While the details (shape, occultation time, etc.) of the light curve of the ionizing source in this model (top panel of Figure 1) may differ from
real sources, the qualitative behavior of the heating-minus-cooling function
is always the same. Moreover, whatever the real light curve of the binary
system is it can always be represented by a sum of modes of oscillation
(Fourier components) and the heating-minus-cooling function will respond
also in the form of a superposition of plane waves, as stated by Equation~(7).
Thus, our formalist is generally applicable.

The time-dependent simulations show phase shifts in the heating-cooling wave for different radii in the nebula. This shift is
determined by the velocity of the radiation/heating fronts across the nebula, which is of the order $0.1\times c$ (see Garc\'{\i}a et al 2013).
Thus, radiation/heating fronts are highly supersonic and that justifies our initial assumption that the gas density remains essentially constant
despite the thermal waves.

An interesting observation from Fig.~1 is that variations in $L$ change 
with depth inside the nebula. While short duration eclipse events lead to 
pulse-like variations in $L$ near the illuminated side of the nebula, these
become wider and decrease in amplitude with increasing depth. Moreover,
deep enough in the nebula (beyond $\sim 2\times 10^{16}$~cm in the simulation of
the amplitude of the  simulation of Fig. 1) the variations blend with each 
other
resulting in a sinusoidal $L$ function around the steady-state case.

\begin{figure}
\rotatebox{0}{\resizebox{\hsize}{\hsize}
{\plotone{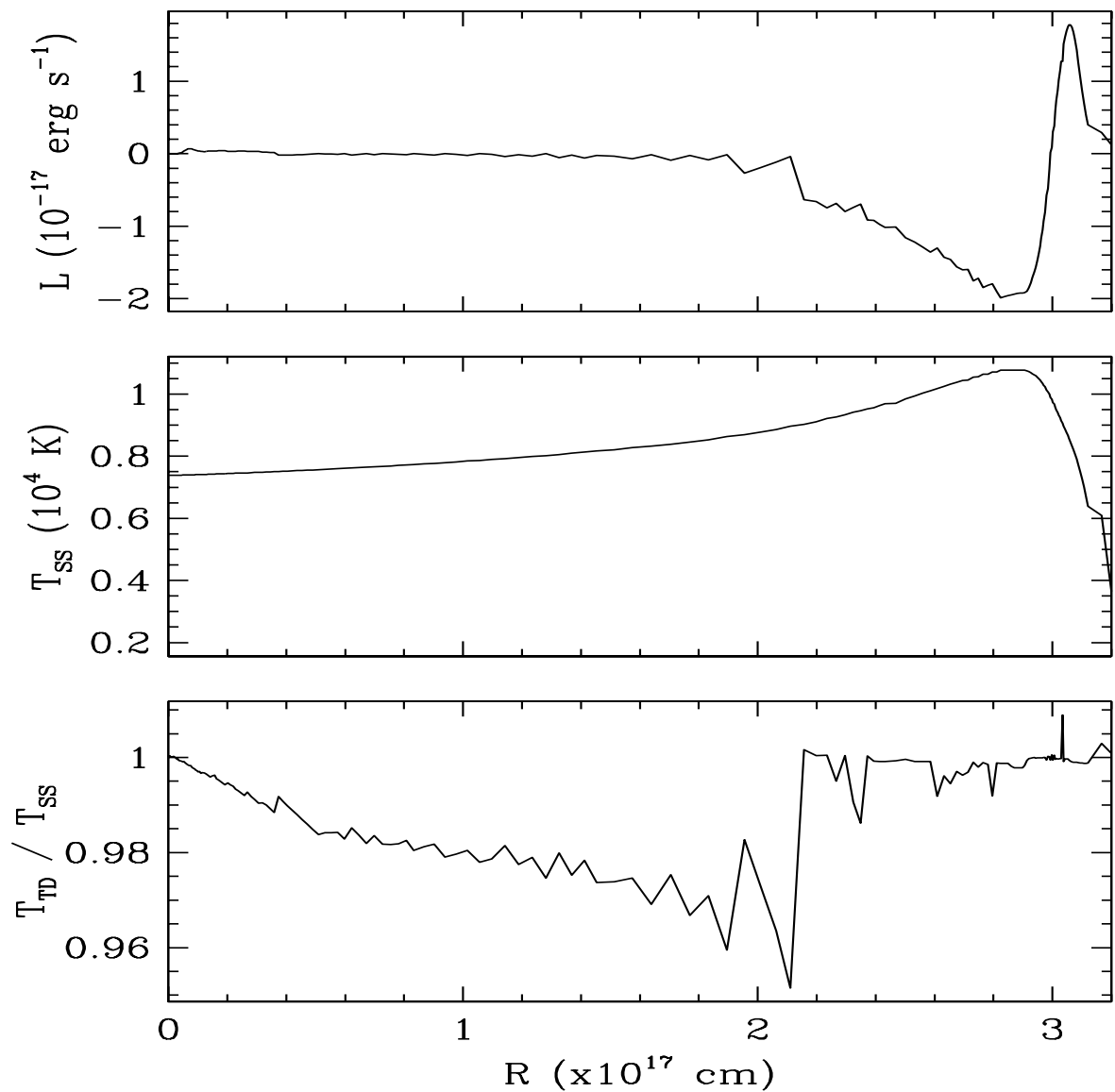}}}
\caption{Heating-minus-cooling, steady state equilibrium temperature ($T_{SS}$),
and ratio of time dependent (TD) temperature to steady state temperature
($T_{TD}/T_{SS}$) vs. radius. 
The simulation conditions are the same as in Figure~1.
}
\label{hmcfunction}
\end{figure}

Figure 2 plots the $L$ function vs. radius in the nebula.
This is from the same simulation as in Figure 1.
It can be seen that cooling tends to dominate over heating at large radii.
This behavior results from the fact that in photoionized plasmas the electron temperature rises
toward the ionization front, as the ionizing radiation field becomes harder. Now, a time variable 
ionizing radiation leads to an ionization parameter that fluctuates between maximum and minimum values.
This in turn, tends to yield a double-peaked  temperature profile
in the nebula
Thus, the overall thermal structure of the nebula will differ from that of the
steady-state cloud (see Figure 2 lower panels).
Notice that, while the new temperatures in this simulation are only a little lower than 
steady-state temperatures, this simulation does not take into account RTF.

Each mode of temperature oscillation in the nebula will satisfy the equation~(6),
thus let us consider one of these modes with
\begin{equation}
\frac{dL}{dt}=i L_1 \omega \exp{i(\omega t - k_1 x)}.
\end{equation}
For this mode, Equation~(6) has solution
\begin{equation}
T_1(x,t)=\frac{L_1\omega}{(k_1v_s)^2-\omega^2} \exp{i(\omega t - \kappa_1 x)}. 
\end{equation}
Thus, thermal perturbations with $k_1\approx \omega/v_s$ are resonantly amplified.

The general solutions to Eqn. (6) is
\begin{equation}
T_1(x,t)=\sum_j \frac{L_j\omega}{(k_jv_s)^2-\omega^2} \exp{i(\omega t - \kappa_j x)}.
\end{equation}
This summation is dominated by the resonant models with $k_j\approx \omega/v_s$.
These are RTF driven by the variability of the ionizing source propagating through the nebula.

\subsection{Propagation perpendicular to the disk}

We have shown that RTF are expected to exist
across the orbital disk of the binary system. The disk is expected to be only a few degrees wide. In this section we study how temperature fluctuation
propagate vertically through the rest of the nebula. We argue that vertical 
propagation happens in two stages. First, thermal waves are powered by the
RTF along the disk. Then, these waves lead to shocks that convert mechanical 
energy into thermal energy and travel through the nebula.

In the direction perpendicular to the orbital disk the linearized perturbation equations are
\begin{eqnarray}
\frac{\partial n_1}{\partial t} = -n_0 \frac{\partial u_1}{\partial y},\\ 
\frac{\partial u_1}{\partial t} = \frac{1}{m} \left(\frac{T_0}{n_0}\frac{\partial n_1}{\partial y}
+ \frac{\partial T_1}{\partial y}\right), \\
3\frac{\partial T_1}{\partial t} = \frac{2 T_0}{m}\frac{\partial u_1}{\partial y}+ L,
\end{eqnarray} 
with the boundary condition
\begin{equation}
T_1(y=0,t) = T_{rtf} cos (\omega t).
\end{equation} 

Figure 3 shows the heating-minus-cooling function as a function of temperature away from the equilibrium temperature. It
can be seen that $L$ is roughly linear with the temperature perturbation $T_1$. From this, we derive the approximate relation
\begin{equation}
L(T_1) \approx -\alpha \times k_B T_1 = - \frac{2\pi}{\tau} \times k_B T_1
\end{equation}
with $\alpha = 8.2\times 10^{-6}$~s$^{-1}$ and $\tau$ is the temperature equilibration time ($\approx 8.9$~days). 
This equation indicates that temperature fluctuations of the order of 1000~K 
will be effectively damped for oscillation periods longer than 8.9~days, and
no resonant fluctuations should be expected. By contrast, short-period binaries have ionizing flux variations along the orbital disk that are too
fast for effective radiative dissipation, thus these fluctuations may grow 
and propagate.

\begin{figure}
\rotatebox{-90}{\resizebox{\hsize}{\hsize}
{\plotone{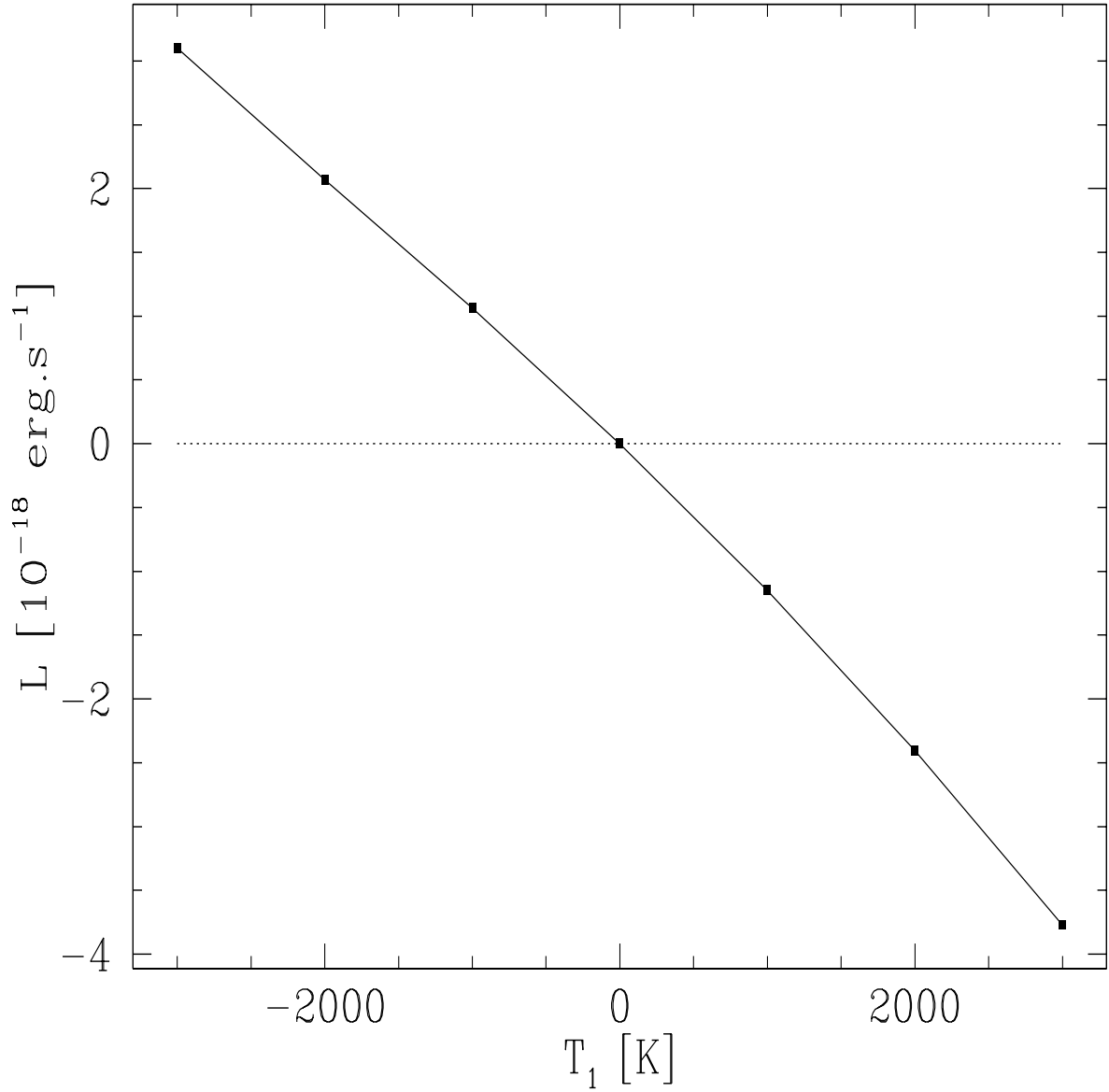}}}
\caption{Heating-minus-cooling vs. perturbation temperature
for a nebula with the same physical conditions as in Figure 1. 
}
\label{hmcvst}
\end{figure}

For temperature, density, and gas velocity perturbations of the form
\begin{eqnarray} 
T_1 = T_{rtf} \exp{[i(\omega t - \kappa y)]}, \\ 
n_1 = N \exp{[i(\omega t - \kappa y)]}, \\
u_1 = U \exp{[i(\omega t - \kappa y)]}, 
\end{eqnarray} 
the conservation equations lead to the dispersion relation
\begin{equation}
\kappa^2 = \frac{\omega^2}{3 v_s^2}\frac{3+(\alpha/\omega)^2+ 2i (\alpha/\omega)}{1+(\alpha/\omega)^2}. 
\end{equation}
In absence of the heating-minus-cooling, $L$, function (i.e. $\alpha=0$) the 
dispersion relation is simply $\kappa^2=\omega^2/v_s^2$. 
Let us consider the case 
$(\alpha/\omega)^2<< 1$, 
i.e. fluctuations with periods significantly shorter than the equilibration time. In this case the wavenumber is 
\begin{equation} 
\kappa \approx \omega/v_s + i/R_1. 
\end{equation} 

The temperature perturbation is now
\begin{equation}
T_1 = T_{rtf} \exp{[i \omega (t - y/v_s)]}\exp{(-y/R_1)} 
\end{equation}
with the attenuation length 
\begin{equation}
R_1\approx {3}\frac{v_s}{\alpha}=\frac{3}{2\pi \sqrt{2}}v_s \tau
\approx 3\times 10^{11}~cm.
\end{equation} 
Thus, the attenuation length is longer than the oscillation wavelength by
a factor $\sim 3\times \omega/\alpha = 3 \tau/Period$. In the case of a short period binary 
star with an occultation period of 3~days the attenuation length is nine 
times longer than the oscillation's wavelength.
In UU~Sagittae the attenuation length is about 60 times longer than the 
perturbation's wavelength.

A key assumption in the treatment above is that the sound speed is constant 
throughout the gas and equal to that of the unperturbed medium. In reality, as
temperature oscillations form the hotter gas travels faster than the cooler 
gas ahead and overtakes it. Thus, it is inevitable that temperature oscillations 
of finite amplitude will steepen into shocks, which convert bulk kinetic energy
into thermal energy. These are expected to be relatively weak shocks, with
Mach numbers not greater than $\sim 2$. They are also radiative shocks, 
with cooling times longer than the binary occultation period but shorter than
the dissipation time scale for compression regions.
Thus, denser gas ahead of the shock front cools down to a new 
equilibrium temperature such that $L=0$. This new equilibrium temperature may 
be somewhat lower than that of the initial state.

Two important conclusions of this section are:

\noindent{(1)} The binary period is a key parameter in the propagation of temperature
fluctuations perpendicular to the
orbital disk. In this regard, the binary period needs to be significantly
shorter than the gas thermalization time. Thus, only short period binary 
stars are expected to exhibit the temperature fluctuations described here.

\noindent{(2)} Temperature fluctuations that propagate through the nebula
are likely to be accompanied by density fluctuations.
At the shock fronts both temperature and density are expected to rise, 
while ahead of the shock fronts the denser gas may cool radiatively.
Thus, we expect a complex structure of temperature and density fluctuations,
possibly out of phase with each other.
This point is demonstrated by spectroscopic observations, as discussed in the next section.

\section{Spectral analysis}

\subsection{Diagnostics of equilibrium temperatures and RTF}

We now turn our attention to the spectra of PNe. In particular, we look at
temperature sensitive line ratios of 
collisionally excited [\ion{O}{3}] and recombination \ion{O}{2} lines.
We consider three different line ratio temperature diagnostics:

\noindent{$R_1$:} \ion{O}{2} $\lambda 4089.29$/\ion{O}{2} $\lambda 4649.13$. This ratio of
two recombination lines was pointed out by Storey et al. (2017) for it is mainly
temperature sensitive while mostly unaffected by density.

\noindent{$R_2$:} $log_{10}($\ion{O}{2} $\lambda 4649.13$/[\ion{O}{3}] $\lambda\lambda4959+5007)$. 
This ratio of a recombination line and a collisional line is temperature sensitive, but
mostly independent of density.

\noindent{$R_3$:} [\ion{O}{3}] $\lambda 4363$/([\ion{O}{3}] $\lambda\lambda4959+5007$). This is a 
classic temperature diagnostic ratio involving only collisionally excited lines.

Figures 4 and 5 plot the line ratios vs. equilibrium temperature $T_0$ for
various values of $T_{rtf}$. In computing these line ratios, the emissivity of
each line is averaged over a temperature fluctuation of the form
\begin{equation}
T(t) = T_0\times (1 + (T_{rtf}/T_0) cos (\omega t)).
\end{equation}
The ratios on the left-hand panels of the figures correspond to the case
of temperature independent density. The plots in the right-hand panels 
correspond to the case
\begin{equation}
n_e (T) = n_e(T_0)\times (T_0/T).
\end{equation}\label{density}
This depicts a case where temperature and density fluctuations evolve 
to a gas pressure equilibrium state.
This seems to be the preferred state of the nebulae with extreme ADF
analyzed below. 

The line ratios in Figures 4  and 5 correspond to $n_e(T_0)=10^3$~cm$^{-3}$
and $5\times 10^3$~cm$^{-3}$, respectively. 
In computing the line ratios, we use Case B line recombination emissivities from 
Storey et al. (2017) and electron impact collision strengths and
A-values from Mendoza and Bautista (2014).
In regards to observed spectra, we study the objects listed by 
Jones et al. (2016) and Wesson et al. (2018, private communication), which include the known PNe with the highest ADF
and PNe with known close binary central stars.
It is clear that the present analysis could explain essentially all nebulae
with large or small ADF. Thus, we concentrate here on the objects with
most extreme ADF, which have been difficult to explain by any previous 
model.

The $R_1$ ratio, which involves two recombination lines, is
significantly enhanced by RTF. 
Thus, temperatures would be severely 
underestimated by this diagnostic line ratio if the effects of RTF were 
ignored.
In this sense, PNe with very large ADF typically exhibit $R_1$ ratios
around 0.35 or greater (see Table 1), which in absence of RTF 
would suggest unphysically low temperatures. 
By contrast, $R_3$ ratios, that involve
collisionally excited lines, tend to severely overestimate the temperature
when RTF are unaccounted for.
PNe with very large ADFs typically show $R_3$ ratios of $\sim 3\times 10^{-3}$,
which would indicate temperatures near 9000~K in absence of RTF, but the
true equilibrium temperatures can be as low as 6000~K for $T_{rtf}/T_0\sim 0.9$.

The $R_2$ ratio, combining collisional and recombination lines, is the most 
sensitive to density fluctuations. 
When allowing for density variations to accompany RTF, this ratio varies by
two orders of magnitude for different $T_{rtf}/T_0$ ratios. This is 
not the case in the constant density case.
Most PNe studied here show mutually consistent $R_1$, $R_2$, and $R_3$ line ratios if RTF are
taken into in the constant pressure scenario (Equation 24).

\begin{figure}
\rotatebox{-90}{\resizebox{\hsize}{\hsize}
{\plotone{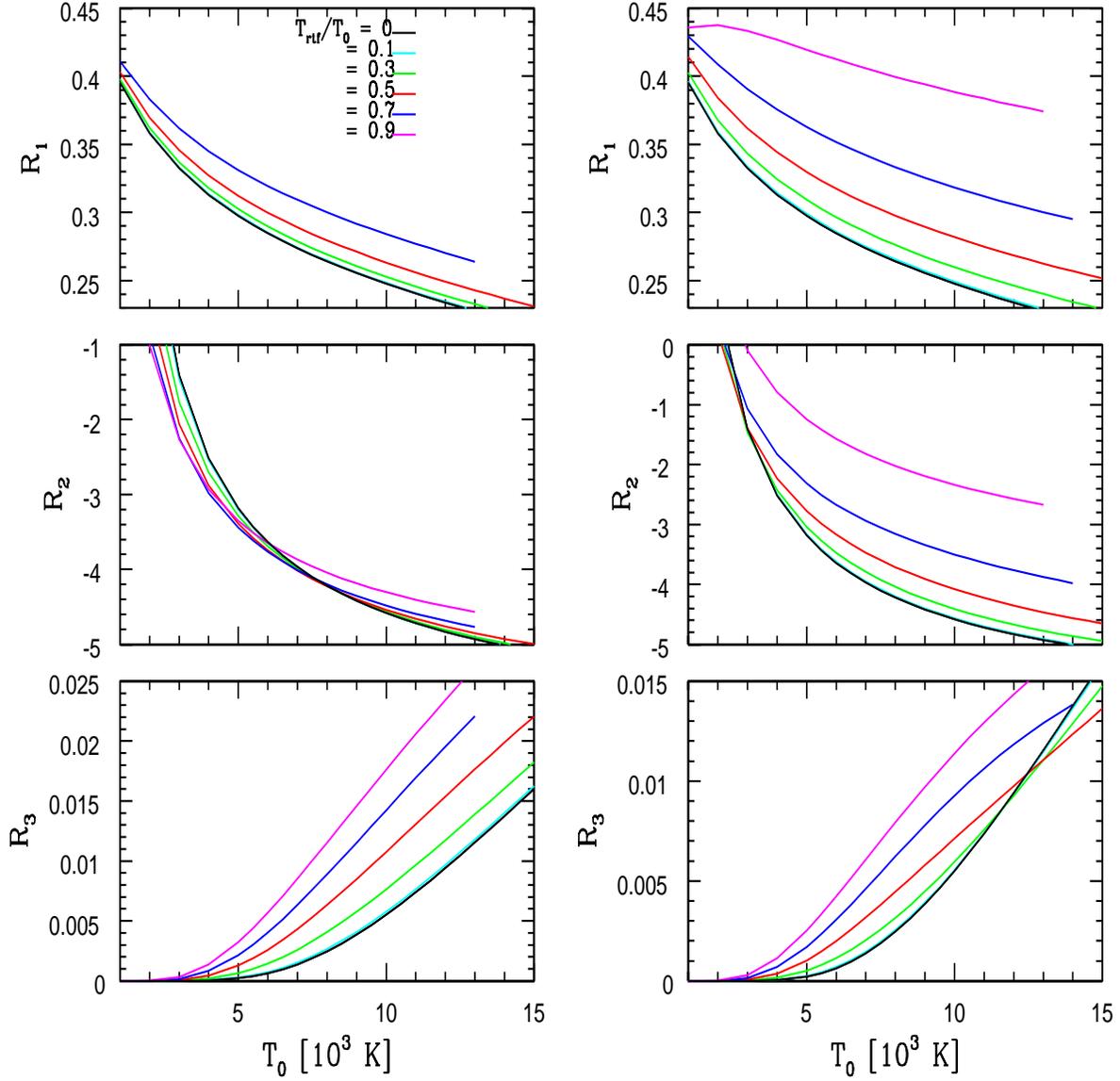}}}
\caption{Temperature line ratio diagnostics of recombination and
collisionally excited oxygen lines vs. nebular equilibrium temperature
for various $T_{rtf}/T_0$ values. Line ratios on the left are for
temperature independent electron density of $10^3$~cm$^{-3}$.
Line ratios on the right are for temperature dependent densities (see text).
}
\label{diagnostics} 
\end{figure}

\begin{figure}
\rotatebox{-90}{\resizebox{\hsize}{\hsize}
{\plotone{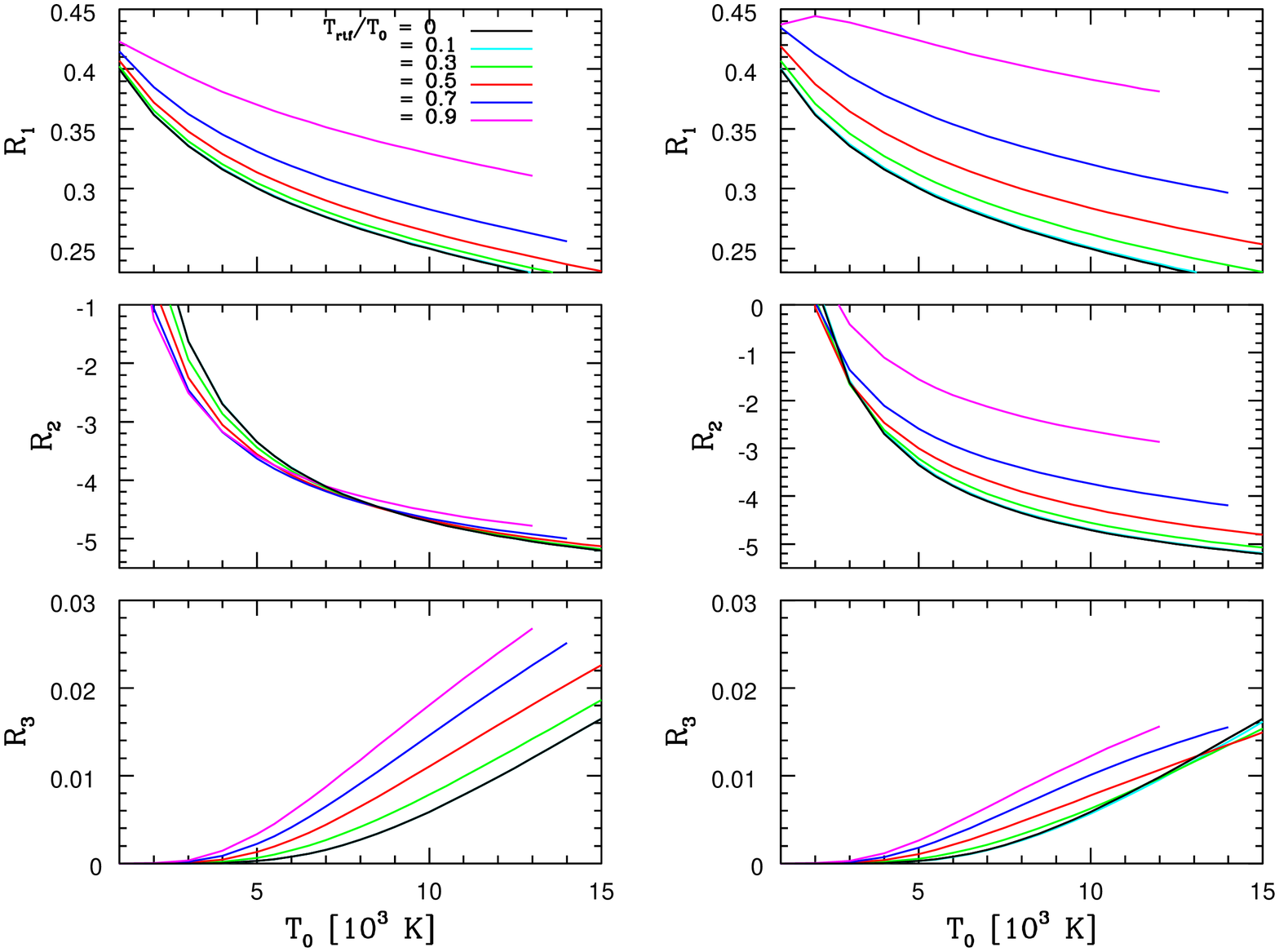}}}
\caption{Like Fig. 4 for electron density of $5\times 10^3$~cm$^{-3}$.
}
\label{diagnostics}
\end{figure}

In Figure 6 we plot the $R_1$ and $R_2$ ratios versus the $R_3$ line ratio.
These are calculated for several $T_{rtf}/T_0$ values from 0 to 0.9 
and electron densities of 1000~cm$^{-3}$ and 5000~cm$^{-3}$.
The left panels of this figure show the case of constant density and the right 
panels depict the temperature dependent density (i.e. constant gas pressure) 
case. In these figures we plot the measured line ratios for the
objects listed in Table~1. We assume uncertainties of 10\% for all line
ratios. 

In the $R_3$ vs. $R_1$ graphs two objects are found outside the range of the plots. These are
A~58 and Fg~1 whose measured $R_1$ ratios are too large, possibly due to 
contamination by unresolved line blends. In addition, there are two objects
within the range of the plots that do not conform to any predicted ratios. These are
Hen~2--283, whose $R_1$ ratio is about three sigma above any theoretical curve, 
 and HH~202 whose measured 
ratio is too low by about a factor of two. Of course, HH~202 in  the Orion nebula, is not a PN

and how the present mechanism for RTF apply to H~II regions is unclear (see discussion in
Section 4). 
Nonetheless, HH~202 is the 
object in Table 1 with the smallest ADF, thus the fact that its measured $R_1$
ratio is smaller than for other objects is consistent with our predictions.
However, we argue that the measure flux for the \ion{O}{2} $\lambda$4089.29 line
by Mesa-Delgado et al. (2009) is underestimated.
In terms of the $R_3$ ratio (involving collisionally excited lines), 
nearly all 
objects 
exhibit measured ratios below 0.011, which is indicative of
equilibrium temperatures lower than $\sim$13,000~K.
The only exception to this is A~30 with $R_3$=0.018, which suggests that either
the equilibrium temperature is unusually large ($\>$15,000~K) or the
measured ratio is overestimated. 

The $R_3$ vs. $R_1$ plots show that the objects with extreme ADF listed in Table~1 
are generally consistent with $T_{rtf}/T_0$ between $\sim$0.7 and $\sim$0.9
and temperature dependent densities (constant gas pressure). By contrast,
models that adopt  constant electron densities generally underestimate the
predicted $R_1$ ratios. 
Thus, it is suggested that temperature fluctuations are accompanied by 
density fluctuations in at least all nebulae with extreme large ADF,
if not in most objects. 

When looking at the $R_2$ vs. $R_1$ plots all objects in Table~1, without exception, are in the plot.
These plots too show that most objects are consistent with temperature dependent density, rather than
constant density. According to these ratios, all objects of Table~1 would have $T_{rtf}/T_0$ ranging from
$\sim$0.3 to $\sim$0.9. Here, the objects with the smallest $T_{rtf}/T_0$ would be 
 A~58, HH~202, and Hen~2--283. HH~202 and Hen~2--283 are the tabulated objects
with the 
smallest reported ADF. On the other hand, A~58 has a large ADF, but the measured fluxes for the 
recombination lines 
appear rather uncertain. 

\begin{figure}
\rotatebox{00}{\resizebox{\hsize}{\hsize}
{\plotone{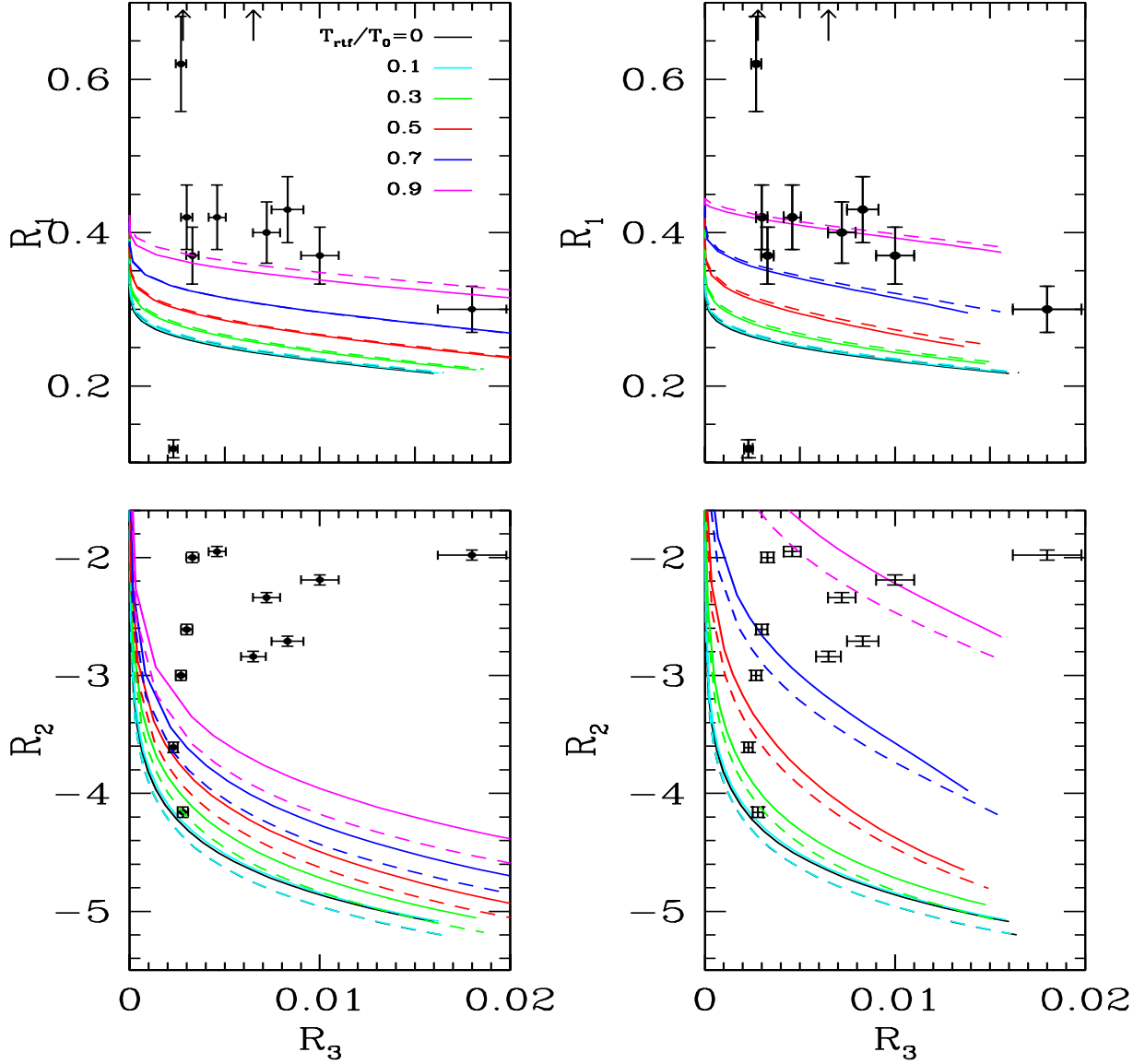}}}
\caption{$R_1$ and $R_2$ line ratios vs. $R_3$. The ratios are shown for $T_{rtf}/T_0$ values 0,
0.1, 0.3, 0.5, 0.7, and 0.9 and for two different electron densities of 1000~cm$^{-3}$ (solid lines) and
5000~cm$^{-3}$ (dashed lines). The left panels depict cases with constant gas density, while the
right panels correspond to the constant pressure scenario. The arrows in the upper-left panel
indicate the position of line ratios for A~58 and Fg~1, whose reported $R_1$ ratios exceed 
the rage of our plots.
}
\label{diagnostics}
\end{figure}

Table 1 presents the line ratios and our diagnosed values for
equilibrium and RTF temperatures for objects listed in 
Jones et al. (2016) and Wesson et al. (2018, private communication).
We note that some uncertainty exists in the line intensities measured in
intermediate spectral resolution, due to unresolved line blends.
For example, the $R_1$ ratio in Fg~1 is unexpectedly large, so we feel it may be
affected by blends.

\begin{deluxetable}{ccccccccl} 
\tabletypesize{\scriptsize}
\tablecaption{Equilibrium and RTF temperature diagnostics of PNe with the
highest known abundance discrepancy factors and other PNe with known
binary central stars}
\tablewidth{0pt}
\tablehead{\colhead{Object} & \colhead{Period (d)} &\colhead{$R_1$}
&\colhead{$R_2$}
&\colhead{$R_3$}
&\colhead{$T_0$}
&\colhead{$T_{rtf}/T_0$}
&\colhead{ADF}
&\colhead{References}}
\startdata
A46 &     0.47 & 0.37 & -2.19 & 0.010 & 7500 & 0.9 & 120 & Corrady et al. (2015)
 \cr
Hf~2--2 & 0.40 & 0.37 & -2.00 & 3.3e-3 & 6500 & 0.8 & 70 & Liu et al. (2006); Schaub et al. (2012) \cr
Hf~2--2 & 0.40 & 0.42      & -1.95 & 4.6e-3 & 6500 & 0.9 & & Wesson et al. (2018) \cr
Ou~5 & 0.36 & 0.42    & -2.61 & 3.0e-3 & 5500 & 0.9$^1$ & 18 & Jones et al. (2016) \cr
A30 & N/A & 0.30      & -1.98 & 0.018 & 10,000 & 0.95$^1$ & $>$100 & Wesson et al. (2003)\cr
A58 & N/A & 1.07$^2$  & -4.16 & 2.8e-3 & 6500 & 0.6 & 89 & Wesson et al. (2008)\cr
HH~202 & N/A & 0.118   & -3.61 & 2.3e-3 & 7000 & 0.4 & 2.2 & Mesa-Delgado et al. (2009) \cr
Fg~1 & N/A & 1.36$^2$ & -2.84 & 6.5e-3 & 6500 & 0.9$^1$ & 80 & Wesson et al. (2018) \cr
Hen~2--283& N/A &0.62 & -3.00 & 2.7e-3 & 6500 & 0.6 & 5 & Wesson et al. (2018) \cr
MPA~1759 & N/A &0.40  & -2.34 & 7.2e-3 & 8500 & 0.8 & 70 & Wesson et al. (2018) \cr
NGC~6337 & N/A &0.43  & -2.71& 8.3e-3 & 9000 & 0.8 & 40 & Wesson et al. (2018) \cr
\enddata
\tablecomments{N/A = Not a confirmed binary.}

\tablecomments{$^1$ More consistent with temperature independent density.}

\tablecomments{$^2$ This ratio seems unusually large. It may require additional measurements}
\end{deluxetable}

\subsection{Computing ionic abundances}

Nebulae with RTF cannot be characterized by a single electron temperature value,
as this may lead to erroneous ionic abundances from collisional and
recombination lines.
Instead, one needs to determine 
equilibrium temperature $T_0$, magnitude of RTF ($T_{rtf}/T_0$),
and the nature of the density structure (constant or temperature-dependent).    
When taking these factors into account the resulting abundances from 
collisional and recombination lines will be into agreement with each other, thus 
resolving the long lasting ADF problem.

Figure~7 shows the emissivities of [\ion{O}{3}]~$\lambda\lambda 4959+5007$,
\ion{O}{2}~$\lambda 4649.13$, and \ion{O}{2}~$\lambda 4089.20$ relative to
H$\beta$. The emissivities are shown for constant
density and temperature-dependent density.

Figure~7 shows that the emissivity of recombination lines relative to H$\beta$
is much less sensitive to the equilibrium
temperature than collisional lines. While the former varies by about a factor of two between 5000 and 
12,000~K, collisional lines change by at least one order of magnitude in the same range of
equilibrium temperatures.
When the plasma density is constant the equilibrium temperature is the most important parameter by far, as
the relative emissivities of collisional and recombination lines are nearly insensitive to 
($T_{rtf}/T_0$). By contrast, under temperature-dependent densities, which seem to prevail in objects 
with large ADF, the relative line emissivities are significantly affected by ($T_{rtf}/T_0$). The effects 
amount to about up to a factor of two between ($T_{rtf}/T_0$)=0 and ($T_{rtf}/T_0$)=0.7 for recombination lines
and about a factor of five for the collisional lines.

Despite the relatively small effects of ($T_{rtf}/T_0$) on the line emissivities with respect to H$\beta$, the
effect of this parameter on abundance determinations must not be underestimated.
Neither recombination nor collisional lines can give reliable abundances in absence of
$T_0$ and $(T_{rtf}/T_0)$. Though, if these temperature conditions are known both
recombination and collisional lines will, in principle, give the same correct ionic abundance. The one remaining source of uncertainty may come from the density structure (constant density vs. constant pressure conditions). In cases where the density structure remained unknown recombination lines should be used in abundance determination.

As an example of how O$^{++}$ abundances are affected by our analysis
we look at the abundances of Hf~2--2. Liu et al. (2006) found that
collisionally excited [\ion{O}{3}] lines are consistent with a temperature of
roughly 9000~K and from this the O$^{++}$/H$^{+}$ fraction is about
$10^{-4}$. By contrast, they find recombination lines consistent
with a temperature of $\sim$900~K and a O$^{++}$/H$^{+}$ fraction of
$\sim 6\times10^{-3}$. When taking RTF into account, a quick
analysis of the spectrum gives $(T_0, T_{rtf}/T_0) \approx (6500 K, 0.8)$
and temperature-dependent density, and from this O$^{++}$/H$^{+}\approx 10^{-3}$.

After determining abundances of the ions present in the spectra one needs to account for ions with no observable signatures. 
Thus, one needs to estimate ionization corrections either by photoionization
modeling or empirical spectroscopic methods. Either way,  this can be 
accomplice with reasonable accuracy if the equilibrium temperature is known
and the abundance of the ions with observable spectral lines is well constrained.
The present approach provides, for the first time, a way to do such determinations.

\begin{figure}
\rotatebox{00}{\resizebox{\hsize}{\hsize}
{\plotone{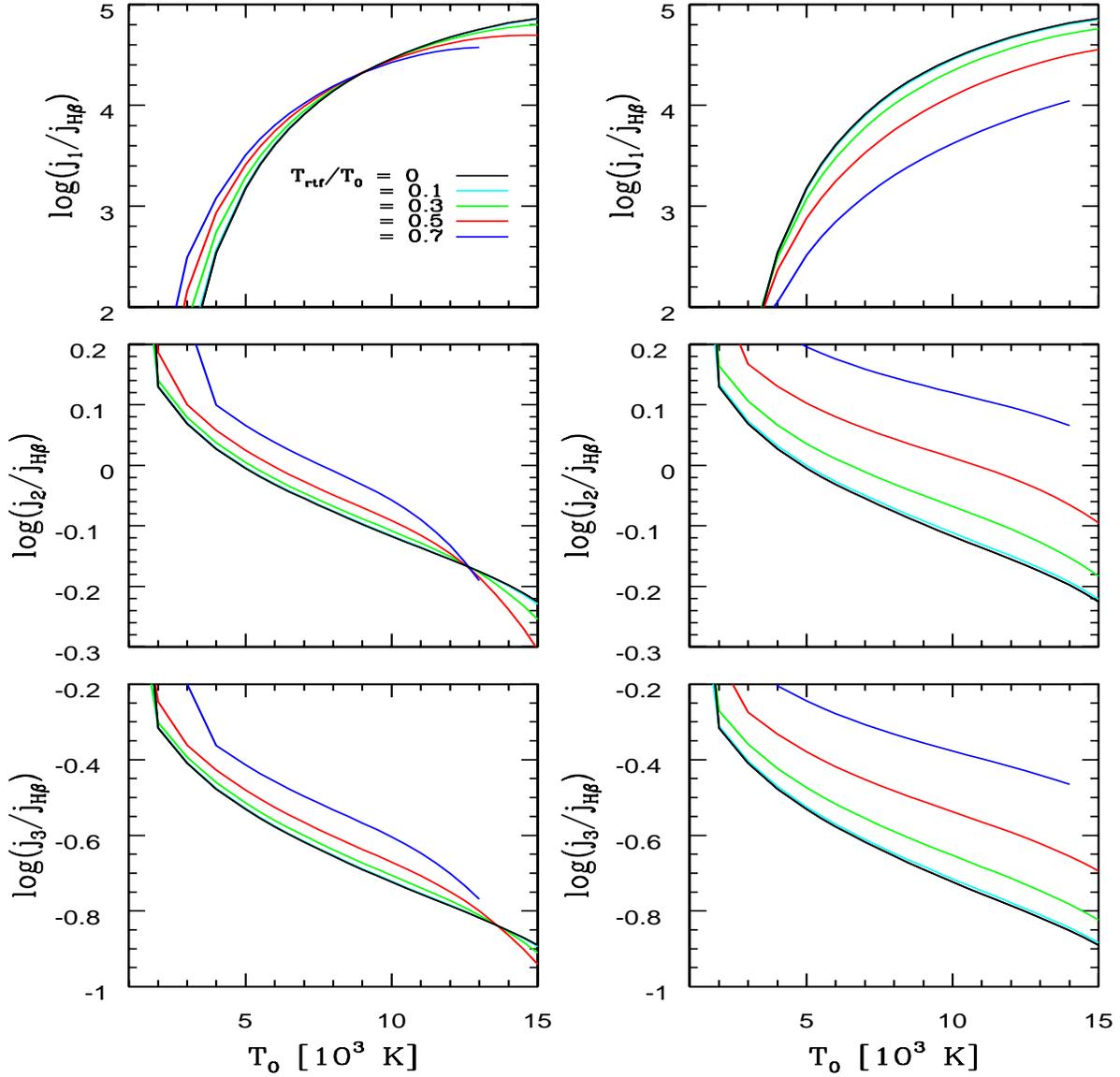}}}
\caption{Line emissivities of the [\ion{O}{3}] $\lambda\lambda4959+5007$
($j_1$), ,
\ion{O}{2}~$\lambda4649.13$ ($j_2$), and \ion{O}{2}~$\lambda4089.29$ 
($j_3$) lines
relative to the emissivity of H$\beta$ vs. equilibrium
temperature. The emissivities are computed for various 
values of $T_{rtf}/T_0$ and temperature independent 
density (left panels) and variable density (right panels).
}
\label{emiss}
\end{figure}

\section{Discussion and Conclusions}

We show that resonant temperature fluctuations, with amplitudes up to 
$\sim$90\% of the equilibrium temperature, are expected to form in PNe photoionized by
short-period-binary stars. Such systems yield a periodically varying ionizing 
radiation field along the orbital disk, which induces
periodic oscillations in the heating-minus-cooling function. As a result, temperature perturbations
in the disk with frequencies similar to those of the ionizing source will undergo resonant amplification.
Further, the temperature fluctuations in the disk cause thermal waves and
shocks that  propagate to the rest of the nebulae. 

Further, our study shows that the amplitude of the RTF depends critically on the
occultation period of the binary star. Only short-period binaries, with 
period of few days, can sustain significant RTF. 

How the present mechanism for RTF applies to H~II regions remains to be studied. On the one hand, H~II regions are 
ionized by one of more young massive stars, which are believed to have a large binarity fraction. Many of these binary
systems could be close binaries, see for instance S~106 (Comer\'on et al 2018). 
On the other hand, if the secondary star is much smaller than the primary it would lead to a very small and thin ecliptic disk,
possibly unable to originate sustainable RTF.

We present diagnostic line ratios that combine [\ion{O}{3}] collisional lines and \ion{O}{2} 
recombination lines. These ratios can be used to estimate the equilibrium temperature and RTF in the O$^{2+}$ region. Similar diagnostics can be created
using different ions that also produce observable collisional and recombination lines
in the nebular spectrum. This will be the subject of future publications. 

When applying these diagnostics to PNe with extremely large ADF and known 
binary central stars we find that they are characterized by equilibrium 
temperatures, $T_0$, between 5500~K and 10,000~K
and $T_{rtf}/T_0$ between 0.6 and 0.9. Most of these objects also show
density fluctuations out of phase with the RTF.

By determining $T_0$ and $T_{rtf}$ one can then estimate the abundance fractions
of different ions relative to hydrogen. These estimate should reflect the
true chemical composition of the nebula by removing the long standing
discrepancies between collisional and recombination lines.

\begin{acknowledgements}
We are grateful to Drs. Jorge Garc\'{\i}a-Rojas, David Jones, and Roger Wesson for useful discussions and comments to the manuscript. Also grateful to
them for making their data available to us for various PNe in advance prior 
to publication.
This work was supported in part by the National Science Foundation (Award AST-1313265). 
\end{acknowledgements}



\end{document}